\documentclass[12pt]{article}

\usepackage[letterpaper,hmargin=1in,vmargin=1in]{geometry}

\usepackage{graphicx,epstopdf,amsmath,amsfonts,amssymb}

\def\be{\begin{equation}}
\def\ee{\end{equation}}
\def\ba{\begin{eqnarray}}
\def\ea{\end{eqnarray}}
\def\ge{\mathrel{\raise.3ex\hbox{$>$\kern-.75em\lower1ex\hbox{$\sim$}}}}
\def\la{\mathrel{\raise.3ex\hbox{$<$\kern-.75em\lower1ex\hbox{$\sim$}}}}

\def\simgt{\mathrel{\raise.3ex\hbox{$>$\kern-.75em\lower1ex\hbox{$\sim$}}}}
\def\simlt{\mathrel{\raise.3ex\hbox{$<$\kern-.75em\lower1ex\hbox{$\sim$}}}}

\newcommand{\bi}[1]{\bibitem{#1}}
\newcommand{\fr}[2]{\frac{#1}{#2}}

\newcommand{\nc}{\newcommand}

\nc{\gone}{\bar g_{\pi NN}^{(1)}}
\nc{\gzero}{\bar g_{\pi NN}^{(0)}}
\nc{\al}{\alpha}
\nc{\ga}{\gamma}
\nc{\de}{\delta}
\nc{\ep}{\epsilon}
\nc{\ze}{\zeta}
\nc{\et}{\eta}
\nc{\ka}{\kappa}
\nc{\rh}{\rho}
\nc{\si}{\sigma}
\nc{\ta}{\tau}
\nc{\up}{\upsilon}
\nc{\ph}{\phi}
\nc{\ch}{\chi}
\nc{\ps}{\psi}
\nc{\om}{\omega}
\nc{\Ga}{\Gamma}
\nc{\De}{\Delta}
\nc{\La}{\Lambda}
\nc{\Si}{\Sigma}
\nc{\Up}{\Upsilon}
\nc{\Ph}{\Phi}
\nc{\Ps}{\Psi}
\nc{\Om}{\Omega}
\nc{\ptl}{\partial}
\nc{\del}{\nabla}
\nc{\ov}{\overline}
\nc{\newcaption}[1]{\centerline{\parbox{15cm}{\caption{#1}}}}
\nc{\us}{U(1)$_S$}

\def\beq{\begin{equation}}
\def\eeq{\end{equation}}
\def\bmat{\begin{displaymath}}
\def\emat{\end{displaymath}}
\def\bear{\begin{eqnarray}}
\def\eear{\end{eqnarray}}
\def\ba{\begin{eqnarray}}
\def\ea{\end{eqnarray}}
\def\bery{\begin{array}}
\def\ery{\end{array}}
\def\bit{\begin{itemize}}
\def\eit{\end{itemize}}
\def\ben{\begin{enumerate}}
\def\een{\end{enumerate}}
\def\btab{\begin{tabular}}
\def\etab{\end{tabular}}
\def\btbl{\begin{table}}
\def\etbl{\end{table}}
\def\bfig{\begin{figure}[htb]}
\def\efig{\end{figure}}
\def\bpic{\begin{picture}}
\def\epic{\end{picture}}

\def\ga{\mathrel{\raise.3ex\hbox{$>$\kern-.75em\lower1ex\hbox{$\sim$}}}}
\def\la{\mathrel{\raise.3ex\hbox{$<$\kern-.75em\lower1ex\hbox{$\sim$}}}}
\def\gappeq{\mathrel{\rlap {\raise.5ex\hbox{$>$}}
{\lower.5ex\hbox{$\sim$}}}}
\def\lappeq{\mathrel{\rlap{\raise.5ex\hbox{$<$}}
{\lower.5ex\hbox{$\sim$}}}}

\def\gyr{{\rm \, G\kern-0.125em yr}}
\def\mev{{\rm \, Me\kern-0.125em V}}
\def\gev{{\rm \, Ge\kern-0.125em V}}
\def\tev{{\rm \, Te\kern-0.125em V}}

\begin{document}

\begin{titlepage}

\setcounter{page}{1}

\vspace*{0.2in}

\begin{center}

\hspace*{-0.6cm}\parbox{17.5cm}{\Large \bf \begin{center}
Direct Detection of Multi-component Secluded WIMPs 
\end{center}}

\vspace*{0.5cm}
\normalsize

\vspace*{0.5cm}
\normalsize

{\bf Brian Batell$^{\,(a)}$, Maxim Pospelov$^{\,(a,b)}$, and Adam Ritz$^{\,(b)}$}

\smallskip
\medskip

$^{\,(a)}${\it Perimeter Institute for Theoretical Physics, Waterloo,
ON, N2J 2W9, Canada}

$^{\,(b)}${\it Department of Physics and Astronomy, University of Victoria, \\
     Victoria, BC, V8P 1A1 Canada}

\smallskip
\end{center}
\vskip0.2in

\centerline{\large\bf Abstract}

Dark matter candidates comprising several sub-states separated by a small mass gap $\Delta m$, and coupled to  
 the Standard Model by (sub-)GeV force carriers, can exhibit non-trivial scattering interactions in  
direct detection experiments. We analyze the secluded \us-mediated WIMP scenario, 
and calculate the elastic and inelastic cross sections for multi-component 
WIMP scattering off nuclei. We find that second-order elastic scattering, mediated by virtual 
excited states, provides strong sensitivity to the parameters of the model for a wide range of mass splittings,
while for small $\De m$  the WIMP excited states have lifetimes exceeding the age of the 
universe, and generically have a fractional relative abundance above 0.1\%. 
This generates even stronger constraints for $\De m \la 200\,$keV due to exothermic de-excitation 
events in detectors.

\vfil
\leftline{March 2009}
    
\end{titlepage}

\subsection*{1. Introduction}

Many concordant aspects of cosmology and astrophysics present us with
compelling evidence for a universe in which dark matter (DM) comprises about one quarter 
of the total energy density in the current Universe.  However, while we have 
ample evidence for its gravitational interaction, the details of any non-gravitational
dynamics if any, and thus its place within particle physics, remains obscure. The
importance of this question, and the strong motivation it implies for physics
beyond the Standard Model (SM), has led to an expansive experimental program, 
both terrestrially and in space, aimed at detecting dark matter through 
non-gravitational interactions, namely
annihilation, scattering or decay \cite{review}.

Among the best motivated scenarios for particle dark matter is a generic 
WIMP - namely a weak-scale massive particle, thermally populated
during the early universe, and subsequently depleted by a weak-scale 
annihilation rate \cite{WIMPS}. This framework has important implications
as it suggests the natural scale for self-annihilation and, albeit in a less direct
manner, gives guidance as to the likely level of scattering with normal matter. 
Consequently, searches for WIMPs present in the galactic halo via their scattering off 
nuclei in radioactively pure underground detectors, have become an integral part of modern 
subatomic physics \cite{cdms,xe,DAMA}. Existing searches have primarily focussed
on nucleon-WIMP elastic scattering, and the null results have placed significant
upper limits on the elastic cross-section that are now beginning to probe the 
natural parameter range \cite{review}.

The chances for successful direct detection depend rather sensitively on whether the WIMP 
is an isolated state, such as real scalar or a Majorana fermion, or a multi-component state, 
such as complex scalar or Dirac fermion. Since the latter allows for the existence of 
vector and/or tensor currents, in many scenarios $Z$- or $\gamma$-mediated 
scattering can result in large elastic cross-sections. However, even a small mass 
splitting $\Delta m$ between the WIMP components -- that we will generically denote $\ch_1$ and $\ch_2$ -- which
allows for coupling to these currents may 
significantly alter the scattering signal if $\De m$ is in excess of the typical kinetic energy of 
the WIMP-nucleus system.
This was first noticed a decade ago \cite{HH,MH}, where it was shown that an otherwise
large sneutrino scattering cross-section off nuclei, mediated by $Z$-exchange, could 
be drastically reduced by splitting the two components of the complex scalar sneutrino by $\Delta m$, 
with $ E_{\rm kin}< \Delta m \ll m_{\rm \tilde\nu}$. This idea was later exploited in Ref.~\cite{WTS} 
where the inelastic scattering of WIMP states split by $\Delta m \sim {\cal O}(100)\,$keV  
was utilized to reconcile the annual modulation of energy deposition seen by DAMA
with the null results of other direct-detection experiments using lighter 
nuclei.
Its of relevance here that the signature of inelastic endothermic WIMP scattering, 
which is enhanced for heavy nuclei \cite{WTS}, differs significantly from elastic scattering. 

The possibility of {\it exothermic} WIMP-nucleus scattering, i.e. with significant energy release,  
was studied in  Ref.~\cite{PRrecomb}, where it was shown that a splitting $\De m \sim {\cal O}(10~{\rm MeV})$
between neutral $\ch_1$ and electrically charged $\ch_2^{\pm}$ components of the WIMP sector will lead to  
WIMP-nucleus recombination with heavy nuclei via the formation of  a bound state between 
the nucleus and the higher-mass negatively charged partner  $\ch_2^-$. A significant amount of energy,
of ${\cal O}(1-10)$ MeV, can be released this way via $e^+$, $\gamma$, $n$ or $\nu$ emission
depending on the particle physics realization of this scenario. 
Ref.~\cite{PRrecomb} also demonstrated the possibility for an ${\cal O}(\Delta m)^{-1}$ enhancement 
in the elastic scattering amplitude due to virtual excitation of the WIMP substructure. 

In general terms, consideration of multi-component WIMP scenarios with relatively small
splittings has in the past been motivated on several fronts. On one hand, due to collider constraints, supersymmetric scenarios 
such as the CMSSM generally require an enhanced annihilation cross-section in the early
universe to avoid over-producing neutralino dark matter, and viable parameter ranges usually
make use of coannihilation with nearby charged states \cite{GS}. Another motivation comes from
the possibility of the catalysis of nuclear reactions in the early Universe leading to the 
resolution of the cosmological lithium problem \cite{cbbn}. MeV-scale splittings also allow for
the possibility of sourcing the galactic 511 keV line via the decay of excited WIMP states \cite{FW,PR511}.

Recently, further motivation for the study of multi-component WIMP scenarios has come
from claims of a positron excess in cosmic rays above 10 GeV  \cite{pamela}, and an
enhancement in the total electron/positron flux around 800 GeV \cite{atic}. While the interpretation
of these anomalies remains an active topic of debate, it is tempting to speculate on  an origin
related to dark matter. However, this requires a significant enhancement of the 
annihilation rate associated with cosmological freeze-out, by factors of 10-1000 depending
on the WIMP mass. This is perhaps most naturally achieved via a new light mediator
with GeV or sub-GeV mass \cite{AFSW,PR}, which allows for a Sommerfeld-type enhancement
of annihilation at low velocities in the present halo. Decays into these metastable mediators
then lead to a dominant leptonic branching fraction for kinematic reasons \cite{AFSW,PR,cholis}.
For the present discussion, the salient feature of the enhancement mechanism
is that the required Coulomb-like potential arises most naturally if the WIMPs are multi-component states, 
such as complex scalars or Dirac fermions, or have at most a small mass splitting between the 
components \cite{AFSW,PR}.\footnote{ 
We note in passing that pseudo-degenerate $\ch_1$ and $\ch_2^\pm$ WIMP multiplets would also boost the 
annihilation cross section by orders of magnitude due to the possibility of ($\ch_2^+\ch^-$) resonant 
scattering, which in particular enhances  neutralino annihilation into gauge bosons and leptons 
provided the neutralino-slepton system is split by $\Delta m \sim $10 MeV \cite{PRrecomb}.} 
   
Perhaps the simplest and most natural realization of this proposal \cite{AFSW,PR} is via a 
secluded \us\ gauge interaction that acts in the dark matter sector 
and couples to SM particles via kinetic mixing with  
SM hypercharge U(1)$_{Y}$ \cite{PRV}. Such a coupling,  $\kappa F^S_{\mu\nu} F^Y_{\mu\nu}$,
provides one of the few renromalizable portals for coupling the SM to new (SM singlet) physics
\cite{holdom}. The particle phenomenology of a (sub-)GeV secluded \us\ sector has 
recently been addressed in a number of studies \cite{pheno}, with the conclusion that the current
sensitivity to $\ka$ is in the range $\kappa \sim {\cal O}(10^{-3}-10^{-2})$. This does not place
 significant restrictions on WIMP properties,  as $\kappa$ does not enter the 
 freeze-out cross section \cite{PRV}, but is interesting in its own right as it is close to
 the natural radiatively generated value \cite{pheno,naturalists}. 

The question of direct WIMP-nucleus scattering has not yet been addressed in detail
for \us-mediated scenarios, despite the high level of recent interest. The cross-section
was calculated to first order in \cite{PRV} in the limit of small $\De m$, where scattering
is mediated by a vector current and the WIMP charge radius. This cross-section is
generically large, and leads to strong limits on $\ka$ \cite{AFSW,PR}, well outside 
both its natural range due to radiative mixing, and consequently the experimental reach 
 \cite{pheno}. While this `secluded' regime is of interest 
for indirect detection, it is important to realize as discussed above that such
a scattering process can be switched off by even a small mass splitting of
the multi-component WIMPs that couple to the \us\ vector current \cite{AFSW,PR}, in 
complete parallel with earlier studies \cite{HH,MH,WTS}. In this case, elastic 
WIMP-nucleus scattering may still proceed, but now at second order with off-shell excited states.  
This echoes earlier calculations in \cite{PRrecomb}, where 
an electrically charged partner of the WIMP state is  virtually produced and 
absorbed in the elastic scattering process. 

In this paper, we calculate the WIMP-nucleus cross sections for inelastic scattering at first order, and
elastic scattering at second order, for the minimal \us-mediated model, and set constraints on combinations of 
parameters, such as the mass splitting $\De m$, the kinetic mixing parameter $\ka$, and 
the mediator mass $m_V$.  We also observe that for light mediator masses and large mixing, the 
interaction may become sufficiently strong that perturbation theory breaks down. Although
Sommerfeld-type enhancement is not important for direct scattering with $\ka \ll 1$ and 
$v\sim 10^{-3}$ in the halo, recombination through the formation of a WIMP-nucleus bound state
can be kinematically accessible for a relevant range of parameters \cite{PR}. 
Another interesting aspect of the minimal \us-mediated WIMP model is the long lifetime of 
$\ch_2$, the excited WIMP state(s), for small $\De m < 2 m_e$. For lifetimes in excess of the age of the universe, 
this immediately raises the possibility of collisional de-excitation within the detector. Not only are such 
processes allowed at first order in perturbation theory, but they will proceed at a constant rate 
even in the limit of very small relative velocity. This can significantly enhance the energy deposition 
within the detector, and we use this signal to set stringent constraints on the 
parameter space of the model. These limits employ an estimate of the minimal
concentration of excited states surviving from the early universe or regenerated locally in the galaxy.

This paper is organized as follows. In the next section we introduce the \us\ model with a mass splitting
of the multi-component WIMPs, its low-energy effective Lagrangian and the lifetime of the excited states. 
Section~3 contains the calculation of elastic and inelastic scattering cross sections for various parameter
choices,  and considers the effects of WIMP-nucleus binding. 
 Section~4 estimates the fractional cosmological freeze-out abundance of excited states,
 their regeneration in the galaxy,  and sets limits on the parameters of the model from de-excitation in
 inelastic exothermic scattering for small $\De m$. We finish with some concluding remarks in Section~5.

\subsection*{2. Multi-component WIMPs with U(1)$_S$-mediators}

The model introduced in Ref.~\cite{PRV} utilizes the kinetic mixing portal to couple the SM and (SM singlet) 
DM sectors via a  \us\ mediator, 
\be
{\cal L} = {\cal L}_{\rm SM} + {\cal L}_{{\rm U(1)}_S} + {\cal L}_{\rm DM}, 
\label{threeL}
\ee
where ${\cal L}_{\rm DM}$ describes the WIMP sector and its \us\ interactions. 
${\cal L}_{{\rm U(1)}_S} $ includes the standard U(1) Lagrangian, the kinetic mixing portal 
and the associated Higgs$'$ sector that gives a mass to the \us\ gauge boson $V_\mu$. After this
symmetry breaking, the relevant part of the low-energy \us\ Lagrangian becomes,
\be
\label{lowV}
{\cal L}_{\rm {\rm U(1)}_S} = -\fr{1}{4}V_{\mu\nu}^2 +\fr{1}{2} m_V^2 V_\mu^2 + 
\kappa V_\nu \partial_\mu F_{\mu\nu},
\ee
where $F_{\mu\nu}$ is the electromagnetic field strength, and a small 
multiplicative shift by $\cos\theta_W$ is absorbed into $\kappa$. 
In this paper, for convenience, we choose the Higgs sector to have twice the \us\ charge of the WIMPs. 
This allows us to introduce two types of (renormalizable) 
mass term in the WIMP action, that itself can be
either bosonic or fermionic:
\ba
{\cal L}^f_{ \rm DM} = \bar \psi (iD_\mu\gamma_\mu - m_\psi) \psi + (\lambda H'\psi\psi + h.c.)~~~~~~~~{\rm fermionic ~DM},
\\
{\cal L}^b_{ \rm DM} = (D_\mu\phi)^* (D_\mu\phi) - m_\phi^2\phi^*\phi+ (\lambda H'\phi\phi + h.c.)~~~~~~~~~{\rm bosonic ~DM}.
\ea
Here $\psi$ (or respectively $\phi$) is the WIMP field in the form of a Dirac fermion (or a charged scalar). 
$D_\mu = \ptl_\mu +ie'V_\mu$ is the usual covariant derivative in terms of the U(1)$'$ gauge 
coupling $e'$, and $\lambda$ is the strength of the dark sector Yukawa interaction. After spontaneous breaking of \us, $H'$ develops a vev, and 
the complex scalar $\ph$ (or respectively Dirac fermion $\ps$), will be split into two real (or Majorana) components which we will
denote collectively as $\chi_1$ and $\ch_2$ with masses $m_i=m_{\ch_i}$. 
The low energy Lagrangian for real scalar WIMPs is then,
\be
\label{lowWb}
{\cal L}^b_{ \rm DM} =  \sum_{i=1,2}\left(\fr12 (\partial_\mu \chi_i)^2 - \fr12 m_i^2 \chi_i^2 \right)
+ e'V_\mu (\chi_1 \partial_\mu \chi_2 - \chi_2 \partial_\mu \chi_1 ),
\ee 
while in the fermionic case, where $\ch_i$ is a two component spinor,
\be
 \label{lowWf}
{\cal L}^f_{ \rm DM} =  \sum_{i=1,2}\left( i\ch_i^\dagger  \bar\si_\mu \ptl_\mu \ch_i -\frac{1}{2}(m\ch_i\ch_i +{\rm h.c.} )  \right)
- ie'V_\mu (\bar\chi_1 \bar\si_\mu \chi_2 - \bar\chi_2 \bar\si_\mu \chi_1 ).
\ee
In both (\ref{lowV}) and (\ref{lowWb},\ref{lowWf}), we have suppressed the details of interactions with the
physical Higgs$'$ as it will not be important for the present discussion.
Of more relevance here is that the WIMP masses are split by an amount that 
scales with the Higgs vev $v'$ and the mass of the mediator:
\be
\label{Deltam}
\Delta m \equiv m_2 - m_1 \simeq  \lambda v'  \sim \fr{\lambda }{e'} m_V \ll m_{1,2},
\ee
where the final inequality reflects the assumed hierarchy, with
the mediator much lighter than the WIMP(s). For the analysis that follows, this is more
relevant than the concrete mechanism for producing 
$\Delta m\sim v'$, as the latter can be achieved within a number of  different model-dependent 
realizations. Eqs.~(\ref{lowV}) and (\ref{lowWb},\ref{lowWf}) constitute the starting point for the calculation 
of $\chi_1$ and $\chi_2$ scattering off nuclei. However, before proceeding in this direction,
 we will first discuss the requirements on the splitting imposed by the need for Sommerfeld-enhanced
 annihilation in the galaxy, and also the lifetime of the excited WIMP states.

 \subsubsection*{2.1 Sommerfeld enhanced annihilation}

Sommerfeld enhancement for the scattering and annihilation of WIMPs
in the galactic halo, with $(\pi \alpha'/v) \gg 1$ when $v\ll \alpha'$ and small $m_V$, 
originates from the modification of free particle wavefunctions due to the impact of the Coulomb
interaction when the particle separation falls below the de Broglie wavelength $(mv)^{-1}$.
Its then clear that this enhancement will only arise (away from possible resonances in
the inter-particle potential) if the mass splitting between the states is smaller than
the Coulomb (or rather Yukawa) potential energy,
\be
\Delta m \la V(r\simeq \lambda_{dB})~~\Longrightarrow ~~ \Delta m \la \alpha' m v \sim E_{\rm kin}\left(\frac{\pi \al'}{v}\right),
\ee
where in the latter relations we also assumed that $m_V \la m v$.  
Therefore, we find that in order to preserve a large enhancement the mass splitting 
must be smaller than the WIMP kinetic energy times the enhancement factor.
Thus, for a TeV-scale WIMP in the galactic halo with $E_{\rm kin}\sim 1$ MeV, requiring an ${\cal O}(10^3)$ enhancement
 to `explain' the PAMELA positron excess \cite{pamela} implies the splitting should not 
exceed 1 GeV. In this paper,  we will consider a range of mass splittings but generally well below this
threshold.

\subsubsection*{2.2 Excited state lifetime}

Once produced, for example in the early universe, the longevity of $\chi_2$ is very important 
due to the possibility for exothermic inelastic scattering in direct detection experiments. 
The decay $\chi_2 \to \chi_1$ depends  sensitively on the size of the 
splitting. For $\Delta m > 2 m_e$, the photon-mediated decay to charged particles dominates, 
and $\chi_2$ decays on time scales much shorter than the age of the Universe
for the range in $\kappa$ that we consider.  Although this may still lead to observable consequences in 
Big Bang Nucleosynthesis, or during later cosmological epochs \cite{FPW}, such effects fall outside 
the scope of the present paper. The other possibility, $\Delta m < 2 m_e$,
only allows decays to photons and neutrinos within the minimal \us\ model. 

The $\chi_2\to \chi_1 \nu\bar\nu$ decay is mediated by $V-Z$ mixing, and its rate summed over three 
neutrino families is calculated to be,
\ba
\label{Ganu}
\Gamma_\nu &=& \fr{4\sin^2\theta_W^4}{315 \pi^3}~\fr{G_F^2\Delta m^9}{m_V^4} ~\fr{\alpha'\kappa^2}{\alpha} \nonumber\\
&\simeq& 3\times 10^{-53}~{\rm GeV} \times \fr{\alpha'}{\alpha}~ \left(\frac{\ka}{10^{-3}}\right)^2 \left(\frac{\De m}{100\,{\rm keV}}\right)^9\left(\frac{100\,{\rm MeV}}{m_V}\right)^4,
\ea
which is much smaller than the inverse age of the Universe, 
$\tau^{-1}_{\rm U} \simeq 1.5\times 10^{-42}$ GeV, unless
$\Delta m > 1$ MeV, and/or $m_V < 10$ MeV. Note that the rate is suppressed 
by the ninth power of the energy released, as compared to the usual $\Delta m^5$ scaling for weak decays. 
This extra suppression can be traced back to the fact that $V-Z$ mixing produces a propagator proportional to $q^2$, 
which is saturated by $\De m$. Similar extra suppression 
factors are to be expected in the decay  rate of any SM-singlet WIMPs with derivative interactions to the SM. 

The decay to photons is mediated by the virtual production of an $e^+e^-$ pair, leading to a
loop-suppressed $ \chi_2\to \chi_1 + 3\gamma$ decay channel. Unlike (\ref{Ganu}), the decay to 
photons involves $V^*\rightarrow 3\gamma$ which is phase-space suppressed but does not require
a weak transition. Assuming a small mass splitting of order 100~keV, we can estimate the rate using 
the calculated $V\rightarrow 3\gamma$ decay width for on-shell $V$ bosons with $m_V\ll 2m_e$ \cite{prv2}, 
\be
\Gamma_V(m_V\ll 2m_e) = \frac{17 \, \alpha^3\al' \kappa^2}{2^7 3^6 5^3 \pi^3} \, \frac{m_V^9}{m_e^8} ,
\label{onshell}
\ee
and replacing $m_V$ by the momentum $q\sim \Delta m$ of the virtual $V$. Since the phase space for the 3-photon decay 
is maximized for momentum $q\sim\De m$, we obtain the estimate,
\ba
 \Ga_{\ch_2 \rightarrow \ch_1 +3\gamma} &\la& \Ga_{V^*\rightarrow 3\gamma} (m_V^*\simeq \De m) 
\times \alpha'\left(\frac{\De m}{m_V}\right)^4 \nonumber\\
&\simeq&  4\times 10^{-47}\,{\rm GeV} \times
\left(\frac{\ka}{10^{-3}}\right)^2 
\left(\frac{\De m}{100\,{\rm keV}}\right)^{13}\left(\frac{100\,{\rm MeV}}{m_V}\right)^4,
\label{3g}
\ea
which is again much smaller than $\tau^{-1}_{\rm U}$, unless $\Delta m$ approaches 1 MeV. Although this
calculation needs to be modified in this limit, we would expect decays above the electron threshold to be fairly
rapid in any case, as noted above.

We conclude that for the parameters of interest here, the excited state lifetime 
can be well in excess of the age of the universe. Thus, a primordial excited state 
population can survive from the Big Bang, while scattering in the halo may also 
re-populate excited states in the late universe. Of course, it is important to emphasize that the 
longevity of the excited states in this model follows from the highly suppressed decay rate 
 to neutrinos and photons. The presence of additional fields, with a mass below 
$\Delta m$, would allow other decay channels and thus a significantly enhanced decay rate. 
For example, one natural possibility within the current scenario would be the presence of 
a very light Higgs$'$ particle, so that $\chi_2 \to \chi_1 + h'$ would be kinematically allowed.
However, in this case the longevity of Higgs$'$ particle may itself have other cosmological and astrophysical 
implications that lead to independent constraints.

\subsection*{3. Elastic and inelastic scattering of multi-component WIMPs}

The basic quantity which determines the mass-normalized counting rate in a given detector is the differential
event rate per unit energy,
\begin{equation}
 \frac{dR}{dE_R} =N_T \frac{\rho_\chi}{m_\chi}\int_{v_{min}}^{v_{max}} d^3v \, v f(v,v_E) \frac{d\sigma}{d E_R}. 
\label{diffrate}
\end{equation}
$N_T$ is the number of scattering centers per unit detector mass, $v n_\chi = v \rho_\chi/m_\chi$ 
is the incident WIMP flux, and $f(v,v_E)$ is the WIMP velocity distribution in the galactic halo
that we take to be Maxwellian with escape velocity in the range $500\,{\rm km}/{\rm s}\, \le v_E \le 600\,$km/s \cite{vesc}. The total number of counts
then follows from integrating over the energy bins, and multiplying by 
the effective exposure, e.g. in kg-days, for a given detector \cite{ls}.

The form of the counting rate (\ref{diffrate}) isolates the physics of the WIMP-nucleon interaction in the
differential cross section $d\si/d E_R$. It has become customary to express the spin-independent differential 
cross section in the  form \cite{review}, 
\begin{equation}
 \frac{d \sigma}{d E_R} = \frac{m_N}{2 v^2} \frac{\sigma_{\rm nucl}^{\rm eff}}{\mu_n^2}
\left[\frac{f_p Z +f_n(A-Z)}{f_{\rm nucl}}\right]^2 F^2(E_R),
\label{sigmanucleon}
\end{equation}
where $m_N$ is the nuclear mass, $\mu_n$ is the WIMP-nucleon reduced mass 
(which is close to $m_p$ for electroweak scale WIMPs),
and $F^2(E_R)$ denotes a possible recoil-energy dependent form factor.
This formula presumes that the force mediating the interaction is short-range, so that the scattering amplitudes 
on individual nucleons $f_{\rm nucl}$ can be considered constant. $\sigma_{\rm nucl}^{\rm eff}$
is then the effective cross section per nucleon, which is now the standard figure of merit 
for comparing the sensitivities of different experiments.

\begin{figure}
\centering
\includegraphics[width=0.29\textwidth]{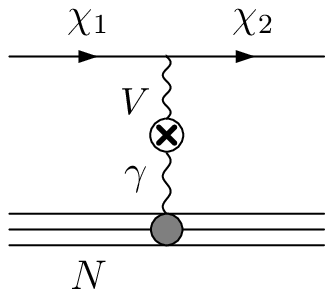} 
\includegraphics[width=0.4\textwidth]{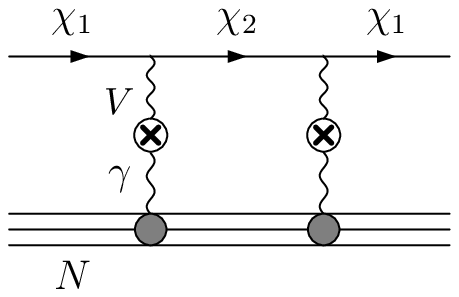}
\caption{First and second Born amplitudes for $\chi_1$-nucleus scattering.}
\label{scatter}
\end{figure}

We now turn specifically to the scattering of quasi-degenerate multi-component WIMPs. There are 
three basic processes that are possible (see Fig.~\ref{scatter}):
\ba
\label{elastic}
&(a)&\mbox{\it elastic scattering}: \;\;\;\;\;\;\,\;\;\;\;\;\;\;\;\;\;\;\;\;\;\;\;\;\;\;\;\;\;\;\;\;\;\;\;\; \chi_{(1,2)} N \rightarrow \chi_{(1,2)} N.\\
\label{endo}
&(b)& \mbox{\it endothermic scattering}\; (Q=-\De m): \;\;\;\;\;\;\, \chi_1 N \rightarrow \chi_2 N.\\
\label{exo} 
&(c)& \mbox{\it exothermic scattering}\; (Q=\De m): \;\;\;\;\;\;\,\;\;\;\;\; \chi_2 N \rightarrow \chi_1 N.
\ea
The $V$-boson vertex in our model always connects $\chi_1$ and $\chi_2$, and thus 
elastic scattering can only occur starting at second order in the Born approximation,
while the inelastic processes (\ref{endo}) and (\ref{exo}) can occur at leading order, and
we will consider these first.

In order to reduce the parameter space of the model when calculating direct detection constraints, we 
make the well-motivated assumption that the ratio $\alpha'/m_\chi$ is restricted to (constant) values which yield the correct 
relic abundance, i.e. for fermionic WIMPs \cite{PRV,PR},
\begin{equation}
 \alpha' = 10^{-2} \times \left(\frac{m_\chi}{270 \,{\rm GeV}} \right).
 \label{abund}
\end{equation}
Further details regarding the WIMP relic abundance are discussed in Sec. 4.

\subsubsection*{3.1 First-order inelastic scattering}

As discussed in the introduction, the relation between $\Delta m$ and the typical kinetic energy of the 
WIMP-nucleus pair is essential. The first-order Born approximation gives the following 
differential WIMP-nucleus scattering cross section: 
\be
\frac{d\sigma_{\chi_{1(2)} \to \chi_{2(1)}}}{d\Om_{f}} = 4\mu_N^2\alpha\alpha'\kappa^2~\fr{|{\bf k}_f|}{|{\bf k}_i|}
\fr{F^2_N(q)}{({\bf q}^2+m_V^2)^2},
\label{first}
\ee
where $\mu_N$ is the reduced mass of the WIMP-nucleus system, ${\bf k}_{i(f)}$ is the initial(final) momentum in the center of mass frame, 
${\bf q}$ is the momentum transfer,  and $F_N(q)$ is the nuclear 
charge form factor, $F_{N}\to Z$ for $q^{-1}\gg r_N$. The magnitude of the momentum transfer 
depends on the kinetic energy in the c.o.m. frame $E_{\rm kin} = {\bf k}_i^2/(2\mu_N)=k^2/(2\mu_N)$, the 
scattering angle $\theta$ and the mass splitting $\Delta m$,
\be
\label{q}
\fr{{\bf q}^2}{2\mu_N} = 2 E_{\rm kin}\left(1-\sqrt{1\mp \fr{\Delta m}{E_{\rm kin}}}
\cos\theta\right)\mp \Delta m,
\ee
where the minus sign corresponds to $\chi_1\to\chi_2$ and the plus sign to $\chi_2\to \chi_1$ scattering. 
The recoil energy of the nucleus in (\ref{diffrate}) is then $E_R = {\bf q}^2/2m_N$. The relation (\ref{q}) illustrates that for the endothermic process (\ref{endo}), $E^{\rm min}_{\rm kin} = \Delta m$, 
while for the exothermic scattering (\ref{exo}), $E^{\rm min}_{\rm kin}=0$ where 
${\bf q}^2(E_{\rm kin}=0) = 2\Delta m \mu_N$. 
In the limit $\Delta m \ll E_{\rm kin}$, Eq.~(\ref{q}) reduces to the standard elastic scattering 
relation, $q=\sqrt{2k^2 (1-\cos \theta)}$.  
In the limit $E_{\rm kin} \mu_N \ll m_V$
and neglecting the 
form factor dependence, $r_N\la (E_{\rm kin}\mu_N)^{1/2} $, 
the total cross section reduces to the elementary formula \cite{PRV}:
\be
\label{simple}
\sigma_{\chi_{1(2)} \to \chi_{2(1)}} =  \fr{16\pi Z^2 \alpha\alpha' \kappa^2 \mu_N^2}{m_V^4},
\ee
which can be interpreted as the scattering of a WIMP due to its finite electromagnetic charge 
radius \cite{ptv}. In the opposite limit of small $m_V$, $E_{\rm kin} \mu_N \gg m_V$, the scattering becomes 
Rutherford-like, and the total cross-section saturates at the minimal momentum 
transfer, $\sigma \sim 1/{\bf q}^2_{\rm min}$, which in an experimental setting would 
be determined by the lowest detectable recoil energy $E_R$.
The generalization of (\ref{simple}) to a finite mass splitting in the limit 
$E_{\rm kin} \mu_N,~\Delta m \mu_N \ll m_V$ is straightforward:
\ba
\sigma_{\chi_{1(2)} \to \chi_{2(1)}} = \fr{16\pi Z^2 \alpha\alpha' \kappa^2 \mu_N^2}{m_V^4}
\sqrt{1\mp \fr{\Delta m}{E_{\rm kin}}}.
\ea
For slow incoming particles with $\Delta m \gg E_{\rm kin}$, we see
that for exothermic scattering the factor $\sqrt{\cdots} \propto 1/v$ so that the rate is 
enhanced by the familiar $\sigma \sim 1/v$ dependence of the cross section. 

\begin{figure}
\centering{
\includegraphics[width=.55\textwidth]{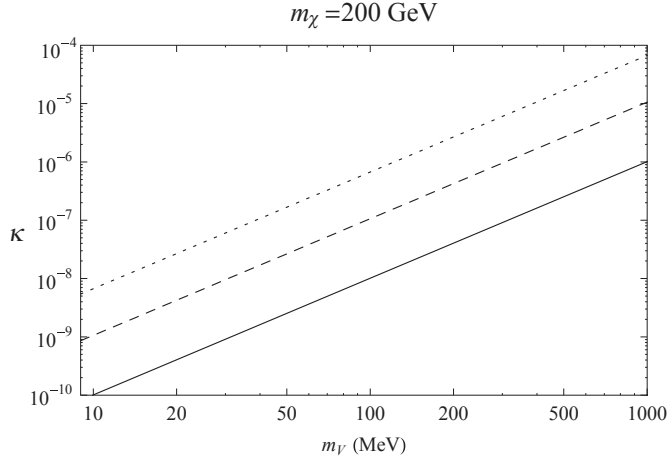}}
\caption{{\it Endothermic inelastic scattering constraints.} 90\% CDMS confidence limits on $\kappa$ as a function of vector mass $m_V$ for a 200 GeV WIMP, with $\alpha'$ chosen to yield the correct thermal relic abundance. We show 
constraints from the endothermic scattering $\chi_1 N\rightarrow \chi_2 N$ for $\Delta m=0$ (solid), 100 keV (dashed), and 150 keV (dotted). No endothermic events are expected in this case for splittings above $\Delta m \sim 190$ keV. }
\label{fig-endo}
\end{figure}

To set constraints on the parameters of the \us\ model, we insert the differential cross
section (\ref{first}) into Eq.~(\ref{diffrate}) and calculate the 90\% Poisson confidence limits (CL) from direct-detection data. In 
order to facilitate as simple a comparison as possible between constraints from endothermic, exothermic, and elastic scattering, in this paper we will restrict ourselves to data from the CDMS experiment \cite{cdms}. Conservative limits result from the assumption that the galactic dark matter component
is dominated by $\chi_1$, and so we focus for the moment on the endothermic first-order process. We 
shall return to constraints following from exothermic $\chi_2 \to \chi_1$ scattering in the next section. 
As can be seen in Fig.~\ref{fig-endo}, the strong sensitivity to $\kappa/m_V^2$ at $\Delta m = 0$ noted in \cite{AFSW,PR} diminishes as $\Delta m $ increases. Moreover, for a given recoil energy, there exists a certain critical value of mass splitting above which kinematics forbids the endothermic inelastic scattering. Thus if $\Delta m$ is large enough  the sensitivity is completely lost, and this occurs for $\Delta m \gtrsim 190$ keV for the parameters used in Fig.~\ref{fig-endo}.

\subsubsection*{3.2 Second-order elastic scattering}

With a larger mass splitting $\Delta m  \gtrsim $ MeV between the 
WIMP $\chi_1$ and its excited partner $\chi_2$, kinematics forbids the first-order scattering process $\chi_1 N\rightarrow \ch_2 N$. However, 
elastic scattering  (\ref{elastic}) can still proceed
via double $V-\gamma$ exchange as depicted in Fig.~\ref{scatter}. For a light $U(1)_S$ vector mediator 
with $m_V \lesssim 50$ MeV or so, the  WIMP-nucleus interaction range is comparable to or larger than the typical 
nuclear radius $r_N$, leading to saturation of the virtual momenta at $| {\bf q}_{\rm virt}|\sim m_V \la r^{-1}_{N}$ 
and resulting in an ${\cal O}(Z^2)$-scaling of the WIMP-nucleus scattering amplitude. 
This implies that such an amplitude cannot be decomposed into a sum of scattering processes on individual 
nucleons. Consequently,  in this regime Eq.~(\ref{sigmanucleon}) is applicable only  as a
definition of the effective nucleon amplitudes that would lead to the equivalent WIMP-nucleus cross section. 
However, once the range of the mediating force falls below 1 fm and $m_V $ starts approaching a GeV, 
WIMP scattering starts probing the internal nuclear structure,  and ultimately 
the scattering amplitude does become a sum of amplitudes for scattering on individual nucleons, as in (\ref{sigmanucleon}). 
Further increase in $m_V$ above the GeV scale allows the \us\ sector to be integrated out
leading to a WIMP-quark effective Lagrangian, that  can be used to calculate $f_{p(n)}$.

In what follows, we calculate the second order elastic scattering amplitude, keeping $m_V$ under a GeV. 
For light and intermediate-mass $U(1)_S$ mediators the WIMP-nucleus cross section can be straightforwardly calculated 
using a non-relativistic potential scattering treatment in the presence of a nuclear charge form factor. 
For $m_V^{-1}$ well below 1 fm, analogous formulae can easily be derived for  WIMP-nucleon scattering. 
The wavefunction of the WIMP-nucleus system under the influence of a central potential is
$\psi( {\bf x})=e^{i {\bf k}_i \cdot {\bf x}} +f(\theta) e^{i {\bf k}_f \cdot {\bf x}}/r$,
where  $f(\theta)$  is the scattering amplitude. 
The second-order Born amplitude for scattering from a potential $V({\bf x})$ can be written  in general as 
\begin{equation}
f(\theta) = -\frac{\mu_N}{2 \pi} \int d^3 {\bf x} d^3 {\bf x}'  e^{-i {\bf k}_f \cdot {\bf x}} 
V({\bf x})G({\bf x} - {\bf x}') 
V({\bf x}') e^{i {\bf k}_i \cdot {\bf x}'},
\end{equation}
where 
$G=( E_{{\bf k}}-E'_{{\bf p}})^{-1}$ is the nonrelativistic propagator for the $\chi_2$-nucleus 
system, which  in momentum space becomes,
\begin{eqnarray}
G({\bf p})& =& \left[ \left( \sqrt{ m_{\chi_1}^2+{\bf k}^2 }+\sqrt{ m_{N}^2+{\bf k}^2} \right)   
- \left( \sqrt{ m_{\chi_2}^2+{\bf p}^2 }+\sqrt{ m_{N}^2+{\bf p}^2} \right)\right]^{-1}   \nonumber \\
&\simeq& \left[ -\Delta m + \frac{ {\bf k}^2}{2 \mu_N} -\frac{ {\bf p}^2}{2 \mu_N}  \right]^{-1},
\end{eqnarray}
where ${\bf p}$ is the intermediate momentum. For a large enough splitting we can neglect the kinetic energy 
terms in the propagator, and this occurs for $\Delta m\gtrsim E_{\rm kin} \sim 100$ keV for a typical nucleus 
with mass of order $m_N=50$ GeV. Retaining only the constant $(\Delta m)^{-1}$ term in the propagator
allows us to use the completeness relation and remove the integration over intermediate states. 
Thus the amplitude greatly simplifies and is expressed via
the square of central potential $V({\bf x}) = V(r)$:
\begin{equation}
f(q) =
\frac{2 \mu_N}{q \Delta m } \int_0^\infty dr \,r \,V^2(r) \sin{q r}.
\label{amp}
\end{equation}
where $q=\sqrt{2k^2 (1-\cos \theta)}$ and $\theta$ are the momentum transfer and 
the scattering angle in the c.o.m. frame, as before. The expression (\ref{amp}) is 
equivalent to the first-order Born formula for the effective potential $V_{\rm eff} = V^2(r)/\Delta m$.

The potential can be written in general as 
\begin{eqnarray}
V(r)&=& \kappa   \sqrt{\alpha\alpha'} \int \frac{d^3{\bf p} }{(2\pi)^3} \frac{F({\bf p})}{{\bf p}^2 +m_V^2}  e^{i {\bf p}\cdot {\bf x}} \nonumber \\
&=& \frac{\kappa Z \sqrt{\alpha\alpha'} }{2 \pi^2 r}\int_0^\infty dp \, \frac{ (F(p)/Z) p \sin{p r}}{p^2+m_V^2} 
\label{V}
\end{eqnarray}
where  $F(p)/Z$ is the charge-normalized nuclear form factor, with $F(0)/Z=1$. To compute this potential, we must 
specify  $F(p)$. Using a model with a uniform charge distribution
within a sphere of finite radius radius $r_N$, 
gives
\begin{eqnarray}
F(p)= \frac{ 3 j_1{(p r_N)}}{p r_N},
\label{ff1}
\end{eqnarray}
where $j_1$ is a spherical Bessel function.
This simple model of the nucleus allows us to analytically compute the potential in Eq. (\ref{V})\footnote{Note that for the first-order endothermic and exothermic inelastic processes we utilize the more common Helm form factor \cite{helm}, which contains an additional exponential factor $\exp(-s^2 p^2)$ relative to Eq.~(\ref{ff1}).
The parameter $s$ is the nuclear skin depth with a typical value of 1 fm, corresponding to energy scales of order 200 MeV. For the case of coherent elastic WIMP-nucleus scattering considered here, the interaction range is approximately $m_V^{-1} \sim r_N > s $, so that the simple solid sphere model of the nucleus is a reasonable approximation.}:
\be
\label{Vofr}
 V(r) = \frac{3 Z \kappa \sqrt{\alpha \alpha'} }{  m_V^2 r_N^3} f(m_V r_N, m_V r)\;\;\;{\rm where}\;\;
 \begin{cases}
\displaystyle{\left.f(x,\hat{r})\right|_{r<r_N} = 1- (1+x)e^{-x} \frac{\sinh \hat{r}}{\hat{r}}.} \\
\displaystyle{\left.f(x,\hat{r})\right|_{r>r_N} = (x\cosh x - \sinh x)\frac{e^{-\hat{r}}}{\hat{r}}.}
\end{cases}
\ee

This is essentially the `Yukawa potential' of a hard sphere, and it is clear from (\ref{Vofr}) that in the limit $m_V \rightarrow 0$ at finite $r_N$, 
the potential reduces to the usual electrostatic potential created by a spherical charge distribution, while 
for $r_N \to 0$, $V(r) \sim \exp(-m_Vr)/r$. Taking both limits simultaneously reduces 
(\ref{Vofr}) to an ordinary Coulomb potential, in which case the scattering amplitude (\ref{amp}) 
simplifies dramatically to give,
\be
f(q) = \fr{\pi Z^2\alpha \alpha' \kappa^2 \mu_N}{q\Delta m }~~\Longrightarrow~~ 
\sigma_{tot} = \fr{2\pi^3 \mu_N^2 (Z\alpha)^4 \kappa^4 \left( \fr{\alpha'}{\alpha} \right)^2}{k^2(\Delta m)^2}
\ln\left(\fr{1}{r_Nm_V}\right),
\label{smallmV}
\ee
where the final integral over momentum transfer is approximated by a logarithm
given the assumption $m_V \ll q \ll r_N^{-1}$. Because the range of validity for (\ref{smallmV})
is almost non-existent, and the nuclear form factor has been neglected, we will not make use
of it for deriving constraints on the model. However, it has the merit of illustrating the 
$Z^4$-scaling of the cross section that translates into the following $Z$-independent prediction for 
the effective cross section per nucleon
\begin{equation}
\label{best}
 \sigma_{\rm nucl} \approx 10^{-42} \, {\rm cm} \times \left(\frac{\alpha'}{\alpha}\right)^2 
\left( \frac{\kappa}{10^{-4}}\right)^4 \left(\frac{{\rm 1~MeV}}{\Delta m}\right)^2~~{\rm for ~small}~ m_V
\end{equation}
where we have inserted typical values for various parameters such as $Z/A\simeq 0.5$, the characteristic recoil energy, etc. 
We can see that direct detection experiments can potentially have strong sensitivity to $\kappa$ for light mediators, due 
to the resulting long range interaction.  Eq.~(\ref{best}) represents the sensitivity in the `best case', which erodes rather 
rapidly with increasing $m_V$.
 
\begin{figure}
\centering{
\includegraphics[width=.55\textwidth]{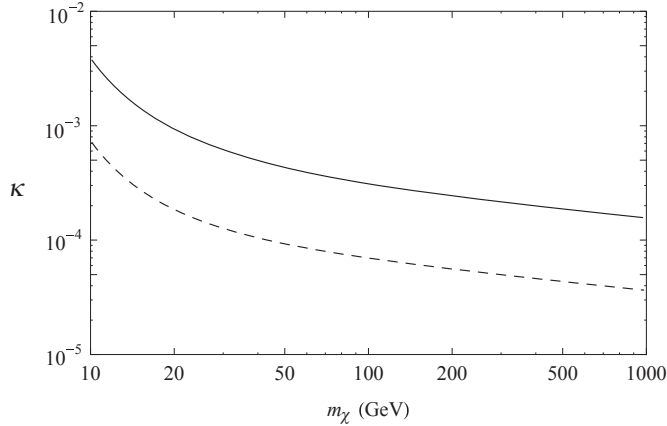}}
\caption{{\it Elastic scattering constraints:} 90\% CDMS confidence limits on $\kappa$ as a function of WIMP mass $m_\chi$ for a mass splitting $\Delta m= 10$ MeV, with $\alpha'$ chosen to yield the correct thermal relic abundance. We show constraints from the elastic scattering $\chi_1 N\rightarrow \chi_1 N$ for $m_V$ = 100 MeV (solid) and 10 MeV (dashed). }
\label{fignuc1}
\end{figure}

To derive actual limits on the parameters of the model,
we write the differential cross section as
\begin{equation}
\frac{d\sigma}{d E_R} = \frac{2\pi m_N }{ \mu^2_N v^2 } |f(E_R)|^2. 
\label{cross}
\end{equation}
Inserting Eq.~(\ref{cross}) with full form factor dependence into (\ref{diffrate}), 
we calculate the constraints on the parameter space using the results from the CDMS 
experiment. These constraints are shown in Fig. \ref{fignuc1} for two cases of 
$m_V =10$ and $100$ MeV. It is easy to see the reduced sensitivity to $\kappa$ as 
$M_V^{-1}$ becomes smaller than the nuclear radius. 

\begin{figure}
\centering{
\includegraphics[width=.55\textwidth]{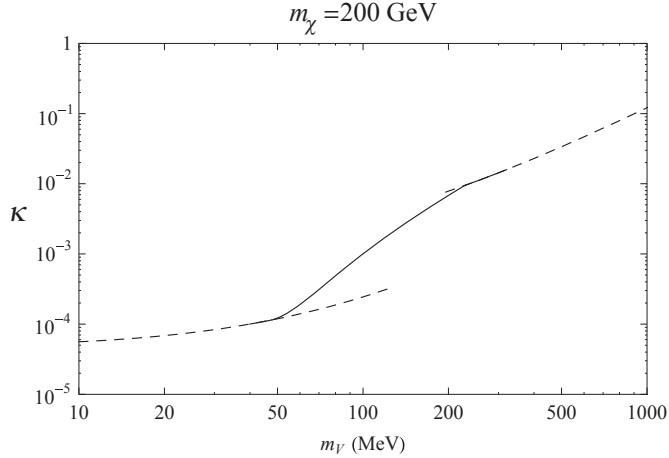}}
\caption{{\it Elastic scattering constraints:} 90\% CDMS confidence limits on $\kappa$ as a function of vector mass $m_V$ for a 200 GeV WIMP with mass splitting $\Delta m =10$ MeV, where $\alpha'$ is chosen to yield the correct thermal relic abundance. We show constraints from elastic scattering $\chi_1 N\rightarrow \chi_1 N$, where the solid line interpolates between the nuclear and nucleon scattering descriptions.}
\label{fignucleus}
\end{figure}

For larger mediator masses,  one can easily observe this change in the $Z$-scaling.
Once $m_V r_N\gg 1$, then  inside the nucleus the potential (\ref{Vofr}) becomes 
$V(r) \sim 3 Z \kappa \sqrt{\alpha \alpha'} m_V^{-2} r_N^{-3}$, and the amplitude at 
small $q$ has the following scaling with $Z$: $f(q=0) \sim Z^2  r_N^{-3} \sim Z$.
The switch from $Z^2$ to $Z$ in the amplitude signifies the loss of coherent nuclear response,
and at that point  the process is best described by scattering on individual nucleons.
Taking $m_V \sim $ GeV, we calculate the WIMP-nucleon elastic scattering amplitude, following the same procedure as developed for the nucleus. We use a proton form factor $F_p(p)=(1+a^2 p^2)^{-2}$ with $a \simeq 0.84$ fm related to the charge radius of the proton.
The amplitude in Eq. (\ref{amp}), the cross section and the rate can all be straightforwardly computed, and we obtain 
the estimate:
\begin{equation}
\label{best2}
 \sigma_{\rm nucl} \approx 10^{-42} \, {\rm cm} \times \left(\frac{\alpha'}{\alpha}\right)^2 
\left( \frac{\kappa}{10^{-1}}\right)^4 
\left(\frac{{\rm 10~MeV}}{\Delta m}\right)^2~~ {\rm for}~m_V\sim 1~{\rm GeV}.
\end{equation}
It is apparent that the loss of coherence in the elastic scattering process means that the sensitivity to 
$\kappa$ in this case is no better than particle physics probes with SM particles. The dependence 
on $m_V$ of the sensitivity is exhibited in Fig.~\ref{fignucleus}, showing the transition
around $m_V\sim 1/r_N$.

Finally, for a fermionic WIMP, a loop-induced effective coupling of the nucleon spin to 
an off-shell photon become possible. This can be interpreted as an effective anapole moment 
of the WIMP  \cite{ptv,PRrecomb}. Although such a coupling will be proportional 
to the first power of $\kappa$, it has no implications for the direct detection, primarily 
because the scattering amplitude is spin-dependent and this  not coherently enhanced, as well as  
being proportional to the small relative velocity of the WIMP-nucleus system.

\subsubsection*{3.3  Non-perturbative scattering and WIMP-nucleus binding}

In the previous section we assumed that the Born approximation is applicable, 
which may not always be the case if $m_V$ is small and $\ka$ is taken to be large.  
For small $\Delta m$ and $m_V$, the criterion for the applicability of the Born approximation 
is $Z\sqrt{\alpha\alpha'} \kappa/v \ll 1$. Since $Z\alpha$ could maximally be of order one, we see that for galactic WIMPs
with $v\sim 10^{-3}$, this is satisfied for any $\ka < 10^{-3}$. We note that for small $m_V$ the exclusion contours in Fig.~\ref{fignucleus}
are well below this threshold and thus well within the range of validity of the Born approximation.
Nonetheless, it is intriguing to consider whether $V-\gamma$ exchange may lead to the formation of 
a bound state as suggested in Ref. \cite{PR}.

For a suitable choice of parameters, it is indeed possible for the 
Yukawa-like potential between the WIMP and a large nucleus 
to support a bound state, or have a quasi-stationary state just above the 
continuum threshold. The potential arises from $V-\gamma$ exchange, 
and in the limit $m_V \ll 1/r_{N}$ takes
the form
\be
 V(r) = - \kappa Z \frac{\sqrt{\al \al'}}{r} e^{-m_v r},
\ee
In the limit of small $\De m$, the existence of at least one bound state then requires (for $m_N \ll m_\ch$)
\be
 \ka \geq 1.4  \frac{m_V}{Z\sqrt{\al \al'}m_N}, \label{thresh}
\ee
which can reach interesting parts of the parameter space for light mediators, and large nuclei. e.g. for the Germanium target at
CDMS, a bound state is possible for $\ka > 10^{-3}$ for $\al' =\al$ and  $m_V \sim 10$~MeV. 

The binding energy is rather small, varying from the Coulomb limit $E_b = (\ka Z)^2\al\al'm_\ch/4$, which could be up to 
${\cal O}(100\,{\rm keV})$ for large $(\ka Z)\sim 0.1$ down to $E_b=0$ at the threshold (\ref{thresh}). 
For the situation where the Dirac WIMP is split by $\De m$ into Majorana components, this potential will inhabit
the off-diagonal terms in a $2\times2$ potential matrix, thus for a mass splitting  $\De m\sim100\,$keV, its apparent
that only large nuclei will allow for recombination.
 
If kinematically accessible, since the recombination rate for $\ch N \rightarrow (\ch N) + \gamma$ is large, we can 
conservatively regard the threshold (\ref{thresh}) as a constraint on the parameters of the model for small 
splittings. This would most likely arise through the formation of too large a relative abundance of 
anomalous heavy elements, but it would also produce a distinctive signature in underground detectors. 
Nonetheless, it is worth remarking that that the strong dependence on the atomic mass means
that the most stringent constraints on the abundance of anomalous heavy isotopes of light elements will, as was the 
case for the pseudo-degenerate scenarios of \cite{PRrecomb}, be less important.
Finally, if  $\Delta m$ is increased or $\kappa$  decreased, the bound state can be pushed into 
the continuum, and may lead to a resonant feature in WIMP-nucleus scattering. 
However, this again will likely be well inside the region excluded by 
Fig.~\ref{fignucleus}.

\subsection*{4. Inelastic scattering and de-excitation constraints}

The presence of an ambient population of excited $\ch_2$ states in the
galactic halo would allow for exothermic inelastic
$\ch_2\to \ch_1$ scattering (\ref{exo}) in detectors. Given the enhanced rate, this can
potentially act as a novel and powerful probe of multi-component WIMP scenarios in which
a sizable excited state population is present. However, these constraints will 
depend sensitively on $\De m$, as a large $O({\rm MeV})$ splitting would result 
in too great an energy deposition within the detector, $E\sim \Delta m m_{ \chi}/(m_{\chi} + m_N)$,
 falling outside the fiducial recoil energy window for most searches. Moreover, larger  
values for $\Delta m$ may considerably shorten the lifetime of $\chi_2$, as in Eq.~(\ref{3g}), 
reducing the local number density. Therefore, in this section, we will concentrate on a relatively small
splitting: $\Delta m \la 500$ keV. 

While the overall abundance of $\chi_2$ and $\chi_1$ is fixed as soon as chemical equilibrium is lost in the
early universe, namely after the temperature drops below $0.05 m_\chi$, the relative 
abundance of $\chi_2$ and $\chi_1$ can vary. The possibility of exothermic scattering in detectors depends 
on the ambient number density $n_2$ of the excited $\chi_2$ population
in the galactic halo, which arises from three main sources, cosmological, galactic, and local:
\be
n_{2} = n_2^{(c)} + n_2^{(g)} + n_2^{(l)}.
\label{3sources}
\ee
The cosmological abundance is regulated by the $\ch_2$  lifetime and 
freeze-out of the $\chi_2\rightarrow \chi_1$ rate in the early universe, while the 
galactic source is related to the inverse possibility for $\chi_1 \to \chi_2$ up-scattering 
in the galaxy, with energy supplied by the WIMP kinetic energy 
\cite{FW,PR511}. The local source could originate from the scattering of $\ch_1$ on
heavy nuclei in the Earth's interior, {\em e.g.} $\chi_1 {\rm Pb} \rightarrow \chi_2 {\rm Pb}$. It is relatively
insensitive to the $\chi_2$ lifetime, provided that it is longer than the WIMP time-of-flight 
through the Earth. However, it is easy to see that the weak scale WIMP-nucleus
cross section cannot lead to more than an ${\cal O}(10^{-8})$ excitation probability.
Thus the cosmological and galactic sources are in general far more important,
but depend crucially on the size of the $\ch_2\leftrightarrow\ch_1$ inter-conversion rate,
which may in general have a number of contributions:
\begin{itemize}
\item {\it Double (de-)excitation}:
\be 
\label{2ways}
\chi_2 + \chi_2 \leftrightarrow \chi_1 + \chi_1.
\ee
This is the most generic possibility, and the only one which will be important in
the minimal \us\ model for a generic range of parameters. Note that it depends
sensitively on the number density of $\ch_2$ states, and thus will tend to
freeze out -- at the latest -- when the temperature drops below $\De m$. This will
allow us to determine the minimal fractional abundance of $\ch_2$ relative
to $\ch_1$ in the \us\ scenario.
\item {\it SM thermalization}:
\be
 \ch_2 + SM \leftrightarrow \ch_1 + SM. \label{chSM}
\ee
For this process to compete with (\ref{2ways}) requires a significant interaction rate between the 
WIMP and SM sectors\footnote{We are grateful to D. Morrissey for pointing out the relevance of SM thermalization.}, and amounts to maintaining thermal contact of the WIMPs with the SM bath 
down to low temperatures $T_\gamma \la \Delta m$. Since we take $\Delta m < m_e$, 
this may only happen if the rate for scattering
off electrons is larger than the weak rate,  $4\pi\ka \sqrt{\al\al'}/m_V^2 > G_F$,
because for $T_\gamma < m_e$ the number density of electrons decreases exponentially. 
However, for the 
fiducial range of values that we consider, $\ka \sqrt{\al\al'}/m_V^2 $ is broadly comparable 
to $G_F$, and the scattering off electrons will have the effect of maintaining
thermal equilibrium between the WIMP and SM sectors  \cite{review, kindec},
possibly all the way to $T_\gamma \sim$ MeV, but will not in itself lead to
depletion of the $\chi_2$ abundance.

\item {\it Dark thermalization}:
\be
\ch_2 + X \leftrightarrow \ch_1 + X.
\ee
This process assumes a more complex dark sector with additional light degrees of freedom.
Given that the dark sector decouples relatively early from the SM bath, and subsequently
cools more quickly, if this additional de-excitation rate can remain in equilibrium well below
this decoupling scale then it could allow for a significant depletion of the excited state fraction.
Such a process then
becomes intertwined with the possibility for $\ch_2 \rightarrow \ch_1 +X$ decay. This is not relevant
for the minimal \us\ scenario for $m_V> {\cal O}$(MeV) as considered here, and so 
we will not consider it in detail. However, it is worth noting that the presence of additional
light states may have further implications for BBN and cosmology, particularly if this interaction
rate is large as would be needed for this process to be important.
\end{itemize}

In the minimal \us\ scenario, for $\Delta m < m_e$, 
the double de-excitation process (\ref{2ways})  is generally the most relevant and we will now consider 
the freeze-out in more detail, in order to estimate the {\em minimal} fractional abundance of $\chi_2$
relevant for direct detection.

\subsubsection*{4.1 Estimate of the fractional abundance of excited WIMPs} 

 In the early universe, chemical freeze-out occurs at relatively high scales, 
where the distinction between $\ch_1$ and $\ch_2$ is unimportant. 
 Since we have $m_\ch\gg m_V$, the primary annihilation process 
 involves $\bar\ch \ch \rightarrow VV$ where 
 the two $V$ bosons are on-shell and subsequently decay to 
the SM via kinetic mixing with the photon. For $m_V \ll m_\ch$ we have \cite{PRV,PR}
\be
 \langle \si v \rangle_{\rm ann} = \frac{\pi (\al')^2}{2m_\ch^2} \rightarrow 2.4 \times 10^{ -26} {\rm cm}^3{\rm s}^{-1}, \label{freezeout}
\ee
where the latter relation follows from ensuring that the WIMPs have 
a relic density that saturates the measured value of $\Omega_{DM}$. 
For our purposes, chemical freeze-out for $\ch_1$ and $\ch_2$ will 
occur at a high temperature scale $T_f \sim m_\ch/20$, and so we can simply
take (\ref{freezeout}) as a constraint relating the coupling $\al'$ to $m_\ch$ as in (\ref{abund}).

If the $\ch_2$  lifetime exceeds $\tau_{\rm U}$, as discussed in Sect.~2.2, and 
 the rate for inter-conversion (\ref{2ways}) is slow -- dropping below the 
Hubble expansion rate before the average WIMP energy falls below $\Delta m$ --
the inevitable prediction is $n_2/n_1 \simeq 1$. 
For example, if in the \us\ model the range of the mediator $m_V^{-1}$ is 
very short (e.g. weak scale), $\chi_2$ and $\chi_1$ will be equally abundant.
However, given a relatively long interaction range for small $m_V$, the 
reaction (\ref{2ways}) could remain in equilibrium down to WIMP energies 
of $E\la \Delta m$, resulting in a significant depletion of $\chi_2$ states.  

 After chemical decoupling, the WIMPs remain in 
thermal equilibrium down to lower temperatures, 
scattering 
off \us\ vectors in the dark sector via the Thomson-like process 
$\ch V \rightarrow \ch V$ and/or off SM charged particles. 
 Scattering off $V$'s becomes
inefficient once the temperature drops below $m_V$, and most of the vector particles decay.
The $\ka$-dependent scattering of WIMPs on SM charged particles can be straightforwardly computed
 e.g. in the temperature interval 1 MeV $\la T \la $ 100 MeV where scattering off electrons is dominant, with a rate scaling 
as $\kappa^2 \alpha \alpha' m_V^{-4} T^5$. 
 After these processes fall below the Hubble rate, the WIMPs are 
 no longer in kinetic equilibrium with the SM thermal bath and begin to cool more rapidly,
maintaining a quasi-thermal spectrum with temperature $T_\ch$:
 \be
 \label{Tchi}
  T_\ch = T_\gamma^2/T_*~~~ {\rm for}~~ T_\gamma < T_*.
 \ee
 For very small $\kappa$, scattering on $V$ dominates, and 
$T_* \sim m_V$. However, if the rate of re-scattering on electrons is comparable
to the weak rate, {\em i.e.} $ 4\pi \kappa (\alpha\alpha')^{1/2}m_V^{-2} 
\sim  10^{-5} m_p^{-2}$,  decoupling is postponed to a temperature $T_*\sim $ MeV.
 
 Once kinetic decoupling from the SM bath is complete, the  inter-conversion process (\ref{2ways}) 
is capable of driving down the fractional abundance of $\chi_2$, provided that it is
faster than the Hubble rate, and the energy 
of the emerging non-thermal $\chi_1$'s is quickly re-distributed in the WIMP sector. 
If such a quasi-thermal state is maintained below $T_\ch < \Delta m$, the 
fractional abundance of excited states will be exponentially suppressed,
 \be
  \frac{n_2}{n_1} \sim \exp\left(-\frac{ \De m}{T_{\ch}}\right). \label{frac}
 \ee
 Due to the finite rate for the 
 $\ch_2\chi_2\to \ch_1\chi_1$ de-excitation process, the approximate 
 freeze-out condition is, 
 \be
 H(T_\gamma^f) = \left[n_2 \langle \sigma_{22\to 11} v \rangle\right]_{T_\chi^f},
 \ee
 where the SM and dark sector freeze-out temperatures are related via 
 (\ref{Tchi}). This results in the following estimate for the
 fractional freeze-out abundance:
 \be
 \label{ratio}
 \fr{n_2}{n_2+n_1} = \eta_b^{-1}~\fr{\Omega_b }{\Omega_{DM }}~\fr{m_\chi}{m_p}~
\fr{H(T_\gamma^f) }{\left[ n_\gamma\langle \sigma_{22\to 11} v \rangle \right]_{T_\chi^f}},
 \ee
 where $\eta_b = 6.2 \times 10^{-10}$ is the baryon-to-photon ratio, 
$\Omega_b/\Omega_{DM} \simeq 0.2$ is the ratio of baryonic to DM energy densities, $m_p$ is the 
proton mass, and $n_\gamma(T) = 0.24 T^3$ is the photon number density. 
The Hubble rate is given by $H(T) \simeq 1.7 g_* M_{\rm Pl}^{-1}T^2$, 
while $M_{\rm Pl} = 1.2\times 10^{19}$ GeV and $g_* \sim 10$. 
 
The minimal value for $n_2/n_1$ is achieved when the $\ch_1\leftrightarrow \ch_2$ inter-conversion cross section is maximized. 
Since we are working in the regime $\alpha'/v \ga 1$ with small $m_V$, perturbation theory is  
not applicable, and the Schrodinger equation should be solved numerically as in the recent paper \cite{Cline}.
However, for our discussion it suffices to saturate $\sigma_{22\to 11}$ by the $s$-wave unitarity
limit or by the range of the force carrier, $m_V^{-2}$, whichever is smaller:
\be
\sigma_{22\to 11}^{max} \sim \fr{\pi}{k^2} ~~{\rm for}~~{k\ga m_V}~~\Longrightarrow~~
\langle \sigma_{22\to 11} v \rangle \la \fr{\pi (T_*)^{1/2}}{m^{3/2}_\chi T_\gamma}.
\label{optimal}
\ee
 Combining (\ref{ratio}) and (\ref{optimal}), we obtain the 
  minimal fractional $\ch_2$ abundance surviving from the early cosmological epoch:
\be
\label{cosmo1}
\left[\fr{n_2}{n_1} \right]^{(c)}_{\rm min} \simeq  10^{-2}\times \left( 
\fr{m_\chi}{300~ {\rm GeV}} \right)^{5/2} \left( \fr{10~ {\rm MeV}  }{T_*}\right)^{1/2},
\ee
where the abundance is also implicitly bounded from above by $[n_2/n_1]_{\rm max}=1$.
Note that if thermalization via electron scattering is negligible, then in order to maximize the rate
one can chose $m_V \sim k \sim ( m_\chi \Delta m)^{1/2}$ at WIMP energies 
comparable to $\Delta m$, so that the optimal 
value for $T_*$ in (\ref{cosmo1}) is $T_*\sim m_V \sim ( m_\chi \Delta m)^{1/2}$. 
More generally, the estimate (\ref{cosmo1}) indicates that for a TeV-scale WIMP, the fractional abundance of 
$\chi_2$ does not drop below 5\%, independent of the strength of the de-excitation reaction. 
We should note that  more accurate computations of the excited fraction $n_2/n_1$ are certainly possible, 
but the estimate (\ref{cosmo1}) will suffice to obtain an estimate of the direct detection rate.

In addition to this cosmological source, there is also the probability of endothermic up-scattering
in the galaxy, adding to the fractional abundance in (\ref{cosmo1}). Since the rates for the forward and backward reactions 
are related, we can estimate the excited state abundance from galactic scattering to be, 
\be
\left[\fr{n_2}{n_1} \right]^{(g)} \sim \tau_{\rm int}\fr{\rho_\chi}{m_\chi} \langle \sigma_{11\to 22} v \rangle_{\rm gal},
\ee
where $\rho_{\chi} \sim 0.3$ GeV cm$^{-3}$ is the local dark matter energy density, $v\sim 10^{-3}$, 
and $\tau_{\rm int}$ the `integration time', equal at least to the age of the Milky Way, 13 bn yr, or the lifetime of 
$\chi_2$, whichever is smaller. Assuming a lifetime in excess of 10 bn yr, we obtain a simple estimate,
\be
\label{naive}
\left[\fr{n_2}{n_1} \right]^{(g)} \sim 10^{-4} \times \left( \fr{300~ {\rm GeV}}{m_\chi} \right)^{3}\times
\exp(-\Delta m/T_{\rm eff}),
\ee
where $T_{\rm eff}$ is the effective WIMP temperature in the galactic halo. This suggests that the 
galactic source is somewhat subdominant, but can in principle compete with the cosmological source
if  the WIMP mass is $\sim 100$ GeV.

\subsubsection*{4.2 Constraints from direct detection}

\begin{figure}
\centering{
\includegraphics[width=.55\textwidth]{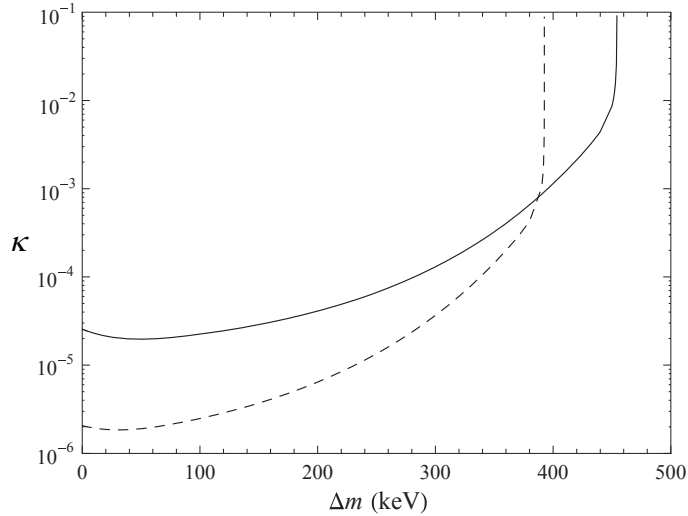}}
\caption{{\it Exothermic inelastic scattering constraints.} 90\% CDMS confidence limits on $\kappa$ as a function of mass splitting $\Delta m$ for a vector mass $m_V=1$ GeV, where $\alpha'$ is chosen to yield the correct thermal relic abundance. We show constraints from exothermic inelastic scattering $\chi_2 N\rightarrow \chi_1 N$ for WIMPs with masses $m_\chi = 100 \,$GeV (solid) and 1 TeV (dashed). The constraints rapidly deteriorate for large $\Delta m$ as in this case most scatterings will have a large nuclear recoil well above the maximum detector sensitivity of $E_R = 100$ keV.}
\label{fig-exotherm}
\end{figure}

Making use of Eq. (\ref{cosmo1}) to determine the minimal fractional $\ch_2$ abundance,
and the results of the previous section,
we are now able to  calculate the exothermic scattering rate as a function of the 
parameters in the \us\ model. 
To be specific, we choose $m_V = 1\,$GeV, and vary $\Delta m$ and $\kappa$, recalling that
elastic scattering processes do not place significant constraints on 
$\kappa$ for such large values of $m_V$. The CDMS constraints for a 100 GeV and  a 1 TeV WIMP are presented in Fig.~\ref{fig-exotherm}. It is clear that for small values of $\Delta m$ exothermic scattering places very stringent 
bounds on values of the mixing parameter $\kappa$. Also, as one expects, the constraints start to deteriorate
for splittings $\Delta m\gtrsim 100$ keV,  as these exothermic events will generally have a
large recoil energy, $E_R \gtrsim 100$ keV, outside the window of most direct detection experiments. 

Another intriguing feature of Fig.~\ref{fig-exotherm} is the increased sensitivity of direct-detection experiments 
for heavy WIMPs. Naively the differential rate (\ref{diffrate}) falls as $m_\chi^{-1}$, but in our case there is a 
hidden $m_\chi$ dependence in the sensitivity coming from both the necessity of larger U(1)$_S$ 
couplings $\alpha'$ to obtain the appropriate cosmological DM abundance, and the WIMP mass 
dependence contained in our estimate of the minimal fractional abundance in Eq. (\ref{cosmo1}).

\begin{figure}
\centering{
\includegraphics[width=1\textwidth]{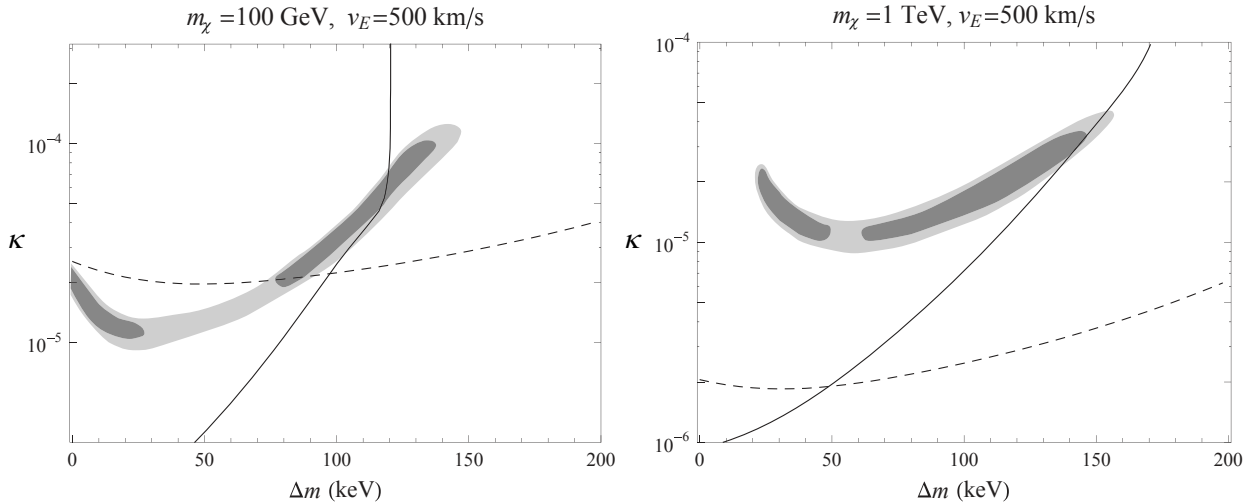}}
\caption{Constraints from endothermic (solid) and exothermic (dashed) scattering in the CDMS detector at 90\% confidence level for a 100 GeV WIMP and 1 GeV U(1)$_S$ vector, with $\alpha'$ chosen to yield the correct thermal relic abundance. We show constraints for a 100 GeV WIMP (left panel) and 1 TeV WIMP (right panel). The complementarity of both constraints is clearly seen. The light (dark) shaded region corresponds to the DAMA 99\% (90\%) CL preferred region, which is excluded by both constraints.}
\label{fig-exo}
\end{figure}

As an application, we consider the feasibility of the minimal U(1)$_S$ scenario to account for the 
DAMA anomaly \cite{DAMA} in light of constraints coming from both endothermic and exothermic scattering. 
The inelastic dark matter scenario \cite{WTS} was proposed to reconcile the null results of direct detection experiments 
with the annual modulation signal observed by the DAMA experiment, and multi-component WIMPs in the U(1)$_S$ 
scenario provide one realization of this idea. To calculate the DAMA preferred regions, we utilize a 
$\chi^2$ goodness-of-fit test \cite{pdg}, including data from the first 12 bins between 
2 keVee and 8 keVee \cite{DAMA}. We show the combined constraints from endo- and exo-thermic processes 
from CDMS in Fig.~\ref{fig-exo} for two values of the WIMP mass, 100 GeV and 1 TeV, along with the 
DAMA preferred regions at the 90$\%$ and 99$\%$ CL. Since the relevant range of $\kappa$, $m_V$ and $\alpha'$ considered 
in this paper implies a rate for WIMP scattering off the thermal electron-positron bath which is close to 
weak-scale even for $\ka\sim 10^{-4}$ as discussed above, the relevant choice of decoupling temperature 
is $T_* \sim {\cal O}(1-10\,{\rm MeV})$.

One observes that in various parameter ranges exothermic scattering can in principle provide more stringent bounds 
on the parameters of the model than endothermic scattering or second-order elastic scattering
for small $\De m$, as shown in Figs.~\ref{fig-endo} and \ref{fignucleus}. One should keep 
in mind, however, the model-dependence of such a conclusion as discussed earlier in this section. Indeed, any 
extension of the minimal \us\ model with additional light particles that could facilitate the decay/de-excitation 
of $\chi_2$ may result in the absence of any significant $\chi_2$ population. The elastic and endothermic 
scattering constraints are, in contrast, completely independent of details concerning the  $\ch_2$ lifetime 
and/or de-excitation rate. 

Finally, it is also worth noting that another possible signature is via ionization due to exothermic WIMP scattering on 
atomic electrons in the detector, which will be essentially monochromatic as it is practically independent of the WIMP velocity,
with a cross section that scales linearly with $Z$.

\subsection*{5. Discussion and conclusions}

In this paper, we have carried out a systematic study of the nuclear scattering of multi-component WIMPs forming part of 
a SM singlet dark sector, focussing on its minimal implementation via a \us\ mediator. The mass splitting of the components
by $\De m$ leads to rich structure for scattering phenomenology. While forbidding first-order elastic  scattering, the
second-order Born cross-section still implies significant sensitivity for direct detection experiments such as CDMS and XENON
for a wide range of mass splittings. Moreover, for small splittings of ${\cal O}(100)\,$keV, first-order inelastic scattering
provides far more stringent constraints, particularly as in this regime the excited states may have a lifetime exceeding the
age of the universe and thus a residual population in the halo can allow exothermic down-scattering in the detector.
This exothermic process leads to additional constraints on specific `inelastic DM' scenarios, via scattering on relatively
light target nuclei that would otherwise have no sensitivity for larger $\De m$.

In rather general terms, it seems there is a clear tension between the constraints discussed here, associated
with enhanced scattering rates, and the presence of relic populations of excited states that may be relevant for other aspects
of  dark matter phenomenology. We will conclude by discussing some specific implications in this vein, but going beyond
the specific \us\ scenario:

\begin{itemize}

\item {\it Inelastic DM and DAMA:} It was noticed a few years ago \cite{WTS}, that a multi-component WIMP scenario with a small
100~keV splitting could resolve the tension between the annual modulation signal observed by DAMA \cite{DAMA}, with the
fact that the required cross sections were seemingly ruled out by the null results of other experiments with lighter materials. 
Despite further null results from XENON \cite{xe}, recent analyses suggest there is seemingly a small parameter range where 
all experiments can be reconciled \cite{inel}. Our simple observation here is that with such a small splitting, the lifetime of a SM 
singlet WIMP excited state will generically be long, and thus the possibility of exothermic scattering of a residual excited state
population will be an important constraint as in Fig.~\ref{fig-exo} for the \us\ scenario. Ameliorating this constraint would require consideration
of models with further decay channels, e.g. via adding a SM charge to the WIMP (as for a sneutrino), or via additional
light states.

\item {\it Multi-component states and the INTEGRAL 511 keV line:} Attempts to attribute the well-measured 511~keV line from the 
galactic center to WIMP interactions generically involve metastable states with decays to positrons. This may be
via delayed decays of a relic \cite{PR511}, or via excitation and decay as in the `exciting dark matter' proposal \cite{FW}.
In the latter case, utilizing the WIMP kinetic energy in the halo requires a scattering cross section well in excess of $s$-wave 
unitarity \cite{PR511,Cline} unless optimistic assumptions are made concerning the galactic halo profile. Recently, it was 
suggested that introducing a three-component WIMP sector $\{\ch_1,\ch_2,\ch_3\}$ \cite{Cline} would resolve this issue by 
utilizing a large relic population of the middle state $\ch_2$, with a small 100~keV splitting from the highest state $\ch_3$. 
Scattering could then excite this transition with a reasonable cross-section, with the subsequent rapid decay to the ground state $\ch_1$  liberating 
${\cal O}(1\,{\rm MeV})$ and sourcing the 511~keV line. In this case its clear that the lifetime and fractional abundance for $\ch_2$ 
must be large, suggesting an analogous problem with the large cross section for exothermic scattering in direct detection. However, it is
worth noting that in this scenario the energy release would be of ${\cal O}(1\,{\rm MeV})$, which may in fact be too large to be easily 
identifiable given the current fiducial recoil energy range.

\item {\it Direct detection of exothermic scattering:} These issues raise the important question of the detectability of exothermic
scattering in direct detection experiments. It is clear that for larger splittings, $\De m > 300\,$keV, the peak of the recoil energy spectrum 
tends to fall outside the conventional fiducial range. It would clearly be of interest to know whether existing experiments are able 
to expand the conventional 10-100~keV fiducial region, in order to provide more stringent constraints on these scenarios.

\end{itemize}

\noindent{\bf Note Added}: While this work was being completed, the preprint \cite{fswy} appeared on the arXiv,
which also considers exothermic scattering through de-excitation, and thus has some overlap with the discussion
in section 4.

\subsection*{Acknowledgements}

We would like to thank J. Cline, D. Morrissey, and I. Yavin for useful correspondence and helpful discussions.
The work of AR and MP is supported in part by NSERC, Canada, and research at the Perimeter Institute
is supported in part by the Government of Canada through NSERC and by the Province of Ontario through MEDT.

\end{document}